\documentclass[12pt,leqno,fleqn]{amsart}
\usepackage{latexsym}
\usepackage{amssymb}
\usepackage{amsxtra}
\usepackage{amsmath}
\usepackage{amsfonts}
\usepackage{CJK}
\def\({\left(}
\def\){\right)}
\def\[{\begin{eqnarray}}
\def\]{\end{eqnarray}}
\def\d{\partial}

\def\d{\partial}

\makeatother       % '@' is restored as a "non-letter" character for TeX

\makeatletter      % '@' is now a normail "letter" for TeX
\@addtoreset{equation}{nosection}
\title{Addition formulae of discrete KP, q-KP and two-component BKP systems }

\author{Xu Gao, Chuanzhong Li\dag,\ Jingsong He\ddag }\dedicatory {
\mbox{}\hspace{-2.5cm}
\ \ \ \ \ Department of Mathematics, Ningbo University, Ningbo, 315211 Zhejiang, P.R.China\\
\dag email:lichuanzhong@nbu.edu.cn\\
\ddag email:hejingsong@nbu.edu.cn}
\thanks{}

\thanks{}

\date{}
\begin{document}

%%%%%%%%%%%%%%%%%%%%%%%%%%%%%%%%%%%%%%%%%%%%%%%%
\begin{abstract}
 In this paper, we constructed the addition formulae for several integrable hierarchies, including the discrete KP, the q-deformed KP, the two-component BKP and the D type Drinfeld-Sokolov hierarchies. With the help of the Hirota bilinear equations and $\tau$ functions of different kinds of KP hierarchies, we prove that these addition formulae are equivalent to these hierarchies. These studies show that the addition formula in the research of the integrable systems has good universality.
\end{abstract}

%%%%%%%%%%%%%%%%%%%%%%%%%%%%%%%%%%%%%%%%%%%%%%%%

\maketitle
PACS numbers: 02.30.Ik, 05.45.Yv\\
Keywords:  the discrete KP hierarchy, the q-deformed KP hierarchy, the two-component BKP hierarchy, D type Drinfeld-Sokolov hierarchy, addition formula, Hirota bilinear equations, $\tau$ function.\\

\allowdisplaybreaks
 \setcounter{section}{0}

%%%%%%%%%%%%%%%%%%%%%%%%%%%%%%%%%%%%%%%%%%%%%%%%%%%%%%%%%%%%%%%%%%%%%%%%%%%%%%%%%%%%%%%%%%%%%%%%%%%%%%%%%%%%%%%%%
%%%%%%%%%%%%%%%%%%%%%%%%%%%%%%%%%%%%%%%%%%%%%%%%%%%%%%%%%%%%%%%%%%%%%%%%%%%%%%%%%%%%%%%%%%%%%%%%%%%%%%%%%%%%%%%%%
%\begin{center}
\section{Introduction}
%\end{center}

The research of soliton and integrable system dates back to the 18th century. The Kadomtsev-Petviashvili(KP) hierarchy as one of the most important subjects contains the following KP equation by promoting the Lax operator:
\begin{equation}
(u_t+6uu_x+u_{xxx})_x+3u_{yy}=0.\\ \nonumber \end{equation}

So far, the series of KP equations have been one of the most important research topics in the field of the classic integrable systems. One important topic in the research of integrable hierarchy is to find the simplest equation to replace the integrable hierarchy, namely the addition formula. The addition formula is also treated as the promotion of the fay identity. The addition formula of the KP hierarchy has been given by generalized Hirota identity in\cite{MS,LVBG1}. We will study the addition formulae for some other special integrable hierarchies in this paper.

Due to the advance of the physics and computer science, the discrete integrable systems have become an extremely essential field in recent years. For example, the discrete lattice system consists of the simple or complex molecules and atomic systems of the physical systems. It is the fact that a growing number of physical problems in the theory and practice are associated with discrete integrable systems. Matrix models of the two-dimensional quantum gravity as an important tool can be closely linked to discrete integrable systems. So is Lorenz group. What's more, we can discretize the continuous integrable system to solve a large number of practical and integrable problems in physics and computer science, which greatly promotes the research of the discrete integrable systems. Therefore, it leads to a series of discrete integrable systems. The discrete integrable system has been widely used in the solid state physics. All these not only show that the discrete integrable systems have important applications and theoretical value, but also promote the study of discrete integrable systems.

In the discrete integrable systems, the research of the discrete KP hierarchy has been the most extensive fields\cite{Kupershimidt}-\cite{L.A}. Obviously, the discrete KP hierarchy has some similar properties with the KP hierarchy\cite{DKJM,dickeybook}, such as tau functions, wave functions, bilinear equations\cite{Iliev,DKJM,dickeybook} and Hamiltonian structure\cite{Kupershimidt,L.A}, etc. Similar to the pseudo differential operator of KP hierarchy, the difference operator $\Delta$ can be used to define the discrete KP hierarchy\cite{Kupershimidt,Iliev,Ghost}. By using $\tau(n;t)$ of the discrete KP hierarchy, we get the form of the Hirota bilinear equations as follows\cite{DKJM,LiuS,LiuSY}:
\begin{equation}
\oint e^{-2\xi(y,z)}\tau(m,x-y-[z^{-1}]) \tau(n,x+y+[z^{-1}]) (1+z)^{m-n}\frac{dz}{2\pi i}=0.\\ \nonumber
\end{equation}
If we set $y=\frac{1}{2}(\sum^{k-1}_{i=1}[\beta_i]-\sum^{k+1}_{i=1}[\alpha_i])$, $[\alpha]=(\alpha,\frac{\alpha^2}{2},\frac{\alpha^3}{3},\cdots),$ instead of expanding in $y=(y_1,y_2,\cdots)$ and computing the integral by taking residues, we get the addition formula for the $\tau$-function of the discrete KP hierarchy
\begin{equation}\label{f11}
\sum^{k+1}_{i=1}(-1)^{i-1}\theta_{\alpha\beta}\tau(n;x+\sum^{k-1}_{j=1}[\beta_j]+[\alpha_i])\tau(m;x+\sum^{k+1}_{j=1,j\neq i}[\alpha_j])(1+\alpha_i^{-1})^{m-n}=0,\end{equation}
which $\theta_{\alpha\beta}$ rests with $\{\alpha_i\}_{1\leq i\leq k+1}$ and $\{\beta_i\}_{1\leq i\leq k-1}$.
Then we can find the fact that the simplest case of the above formula in case of $k=2$ and $m\neq n$ is the Fay identity of the discrete KP hierarchies, so we will give the proof that the discrete KP hierarchies are equivalent to the addition formula.

The q-deformed KP hierarchy is also an important hot topic in the research of the integrable system, which is based on quantum calculus\cite{Klimyk,Kac} and related to the q-deformed quantum integrable models\cite{Bracken}, the q-deformed Bose gas\cite{Shu}, the q-deformed thermodynamics and Virial theorem\cite{Lavagno,ZhangJZ}, etc. The q-KP hierarchy can be viewed as differential equations for the tau function $\tau_q(x;t)$. With the Hirota bilinear equations we get the addition formula:
\begin{equation}\label{f12}
\sum^{k+1}_{i=1}(-1)^{i-1}\theta_{\alpha\beta}\tau_q(x;u+\sum^{k-1}_{j=1}[\beta_j]+[\alpha_i])\tau_q(x;u+\sum^{k+1}_{j=1,j\neq i}[\alpha_j])=0.\end{equation}
Thus eq.$(1.2)$ equals the q-KP hierarchy.

The KP hierarchy as well as the multi-component KP hierarchy are among the most important ingredients of the theory of integrable partial differential equations. They apply in various fields of physics from the hydrodynamics to string theory\cite{LVBG}. The multi-component KP hierarchy with many special sub-series, especially the two-component BKP hierarchy has attracted much attention. The two-component BKP hierarchy is a system with two sets of time variables $t_1,t_3,\cdots, t_{2n+1}$ and $\bar{t}_1,\bar{t}_3,\cdots,\bar{t}_{2n+1}$ with odd indices. The addition formula for the $\tau$-functions $\tau=\tau(t,\bar{t})$ of the two-component BKP hierarchy is:
\begin{align}\label{f13}
\sum^k_{i=1}\prod^k_{j=1,j \neq i}\frac{\alpha_i+\alpha_j}{\alpha_i-\alpha_j}\tau(x+2[\alpha_i]_o,\bar{x})
\tau(x+2\sum^k_{j=1,j \neq i}[\alpha_j]_o,\bar{x}+2\sum^k_{j=1}[\bar{\alpha}_j]_o) =\\ \nonumber
\sum^k_{i=1} \prod^k_{j=1,j \neq i}\frac{\bar{\alpha}_i+\bar{\alpha}_j}{\bar{\alpha}_i-\bar{\alpha}_j}\tau(x,\bar{x}+2[\bar{\alpha}_i]_o) \tau(x+2\sum^k_{j=1}[\alpha_j]_o,\bar{x}+2\sum^k_{j=1,j \neq i}[\bar{\alpha}_j]_o).\end{align}

By reducing the two-component BKP hierarchy, we get the D type Drinfeld-Sokolov hierarchy. So in the end of the paper, we express the addition formula of the D type Drinfeld-Sokolov hierarchy as:
\begin{align}\label{f14}
\sum^k_{i=1}\prod^k_{j=1,j \neq i}\frac{\alpha_i+\alpha_j}{\alpha_i-\alpha_j}\tau_{ds}(x+2[\alpha_i]_o,\bar{x})
\tau_{ds}(x+2\sum^k_{j=1,j \neq i}[\alpha_j]_o,\bar{x}+2\sum^k_{j=1}[\bar{\alpha}_j]_o)(\alpha_i^{-1})^{2nj} =\\ \nonumber
\sum^k_{i=1} \prod^k_{j=1,j \neq i}\frac{\bar{\alpha}_i+\bar{\alpha}_j}{\bar{\alpha}_i-\bar{\alpha}_j}\tau_{ds}(x,\bar{x}+2[\bar{\alpha}_i]_o) \tau_{ds}(x+2\sum^k_{j=1}[\alpha_j]_o,\bar{x}+2\sum^k_{j=1,j \neq i}[\bar{\alpha}_j]_o)(\bar{\alpha}_i^{-1})^{-2j}.\end{align}

In the paper, we show the fact that the discrete KP, the q-KP and the two-component BKP hierarchies are equivalent to the simplest addition formulae. Specially, eqs. (\ref{f11})$\thicksim$(\ref{f14}) are main results of the paper. Our study not only shows the conclusion in the continuous integrable system, but also discusses whether the addition formula will exist in the discrete and multi-component integrable systems. In other words, the addition formula is of good universality.

This paper contains 7 sections. In section 2, we introduce the Hirota bilinear equations of the discrete KP hierarchy and deduce its addition formula. The most important point is to prove the equivalence between the discrete KP hierarchy and its addition formula. In section 3, we introduce the Hirota bilinear equations of the q-KP hierarchy and deduce its addition formula. In section 4, we introduce the Hirota bilinear equations of the two-component BKP hierarchy and deduce its addition formula. In section 5, we study the D type Drinfeld-Sokolov hierarchy. At last, on the one hand, we make a simple conclusions and discussions; on the other hand, we will introduce our future work about the addition formula.

%\begin{center}
\section{The addition formula of the discrete KP hierarchy }
%\end{center}
\setcounter{section}{2}

The discrete KP hierarchy is an interesting object in the research of the integrable systems. Different operators have been used in different
integrable systems. Our purpose is to prove the addition formula of the discrete KP hierarchy. The discrete KP hierarchy can be viewed as the classic KP hierarchy with the continuous derivative $\frac{\partial}{\partial x}$
replaced formally by the discrete derivative $\Delta$ whose action on function $f(n)$ as
\[\Delta f(n)=f(n+1)-f(n).\]

Some basic known facts about the discrete KP hierarchy can be mentioned in Refs.\cite{Iliev,Ghost}. Firstly set a space $F$ be a ring of function, which includes a discrete variable $n\in\mathbb{Z}$ and infinite time variables $t_i\in\mathbb{R}$, namely
\[F=\left\{ f(n)=f(n,t_1,t_2,\cdots,t_j,\cdots);  n\in\mathbb{Z}, t_i\in\mathbb{R}\right\}.\]
The shift operator $\Lambda$ and the difference operator $\Delta$ can be defined as
\[\Lambda f(n)=f(n+1)
\]
and
\[\Delta f(n)=f(n+1)-f(n)=(\Lambda -I)f(n),
\]
where $I$ is the identity operator.
Obviously, the following formula holds true for any  $j\in\mathbb{Z},$
\begin{equation}\Delta^j\circ
f=\sum^{\infty}_{i=0}\binom{j}{i}(\Delta^i
f)(n+j-i)\Delta^{j-i},\hspace{.3cm}
\binom{j}{i}=\frac{j(j-1)\cdots(j-i+1)}{i!},\label{81}
\end{equation}
where the symbol $``\circ"$ represents multiplication.
So we obtain an associative ring $F(\Delta)$ of formal pseudo
difference operators, with the operation $``+"$ and $``\circ"$
\[F(\Delta)=\left\{R(n)=\sum_{j=-\infty}^d f_j(n)\Delta^j, f_j(n)\in
R, n\in\mathbb{Z}\right\},
\]
which $R$ contains two subrings:
$R_+:=\sum_{j=0}^d f_j(n)\circ\Delta^j$
and
$R_-:=\sum_{j=-\infty}^{-1}f_j(n)\circ \Delta^j$.
The adjoint operator $\Delta^*$ can be defined to the $\Delta$ operator:
\[\Delta^* f(n)=(\Lambda^{-1}-I)f(n)=f(n-1)-f(n),
\]
where $\Lambda^{-1} f(n)=f(n-1)$, as well as
\[\Delta^{*j}\circ
f=\sum^{\infty}_{i=0}\binom{j}{i}(\Delta^{*i}f)(n+i-j)\Delta^{*j-i}.
\]
Then we obtain the adjoint ring $F(\Delta^*)$ to the
$F(\Delta)$. For the adjoint operator we have $(PQ)^*=Q^*P^*$ and $f(n)^*=f(n)$.

Being similar to KP hierarchy, the following properties of the wave function and tau function can be given by \cite{Iliev}.

The discrete KP hierarchy is a family of evolution equations depending on infinitely many variables $t=(t_1,t_2,\cdots)$\cite{Iliev,Ghost}:
\begin{equation}
\frac{\partial L}{\partial t_i}=[B_i, L],\ \ \ B_i:=(L^i)_+,
\end{equation}
where $L$ is a general first-order pseudo difference operator
\begin{equation} \label{laxoperatordkp}
L(n)=\Delta + \sum_{j=0}^{\infty} f_j(n)\Delta^{-j}.
\end{equation}
Define the following dressing operator,
\[W(n;t)=1+\sum^\infty_{j=1}w_j(n;t)\Delta^{-j},
\]
which satisfies
\begin{equation}
L=W\circ \Delta\circ W^{-1}.\label{88}
\end{equation}
The Baker-Akhiezer wave function $\Phi(n;t,z)$ and adjoint Baker-Akhiezer wave
function $\Phi^*(n-1;t,z)$ can be used to show the Hirota bilinear identity of the discrete KP hierarchy \cite{Iliev,Ghost,MPV}:
\[\label{Phi}
\Phi(n;t,z)&=&W(n;t)(1+z)^n \exp(\sum^\infty_{i=1}t_i z^i)\\ \notag
        &=&(1+\frac{w_1(n;t)}{z}+\frac{w_2(n;t)}{z^2}+\cdots)(1+z)^n \exp(\sum^\infty_{i=1}t_i
        z^i)\label{41},
\]
and
\[\label{Phi}
\Phi^*(n;t,z)&=&(W^{-1}(n-1;t))^*(1+z)^{-n} \exp(\sum^\infty_{i=1}-t_i
z^i)\\ \notag
&=&(1+\frac{w_1^*(n;t)}{z}+\frac{w_2^*(n;t)}{z^2}+\cdots)(1+z)^{-n}
\exp(\sum^\infty_{i=1}-t_i z^i),
\]
which satisfies
\[
L(n)\Phi(n;t,z)&=&z\Phi(n;t,z),\ \ \ L^*(n)\Phi^*(n;t,z)=z\Phi^*(n;t,z),
\]
and
\[
\d_{t_j}\Phi(n;t,z)&=&B_j(n)\Phi(n;t,z),\ \ \ \d_{t_j}\Phi^*(n;t,z)=-B_j^*(n-1)\Phi^*(n;t,z).
\]
Also one has a tau function $\tau(n;t)$ for the discrete KP hierarchy \cite{Miwa}, which satisfies
\begin{equation}\label{f217}
\Phi(n;t,z)=\frac{\tau(n;t-[z^{-1}])}{\tau(n;t)}(1+z)^n
\exp(\sum^\infty_{i=1}t_i z^i),
\end{equation}
and
\begin{equation}\label{f218}
\Phi^*(n;t,z)=\frac{\tau(n;t+[z^{-1}])}{\tau(n;t)}(1+z)^{-n}
\exp(\sum^\infty_{i=1}-t_i z^i),
\end{equation}
where $[z^{-1}]=(\frac{1}{z},\frac{1}{2z^2},\frac{1}{3z^3},\cdots)$.

From the above, the following theorem can be achieved. The wave function and adjoint wave function satisfy the following bilinear identity. Namely the Hirota bilinear identities of the discrete KP hierarchy are\cite{Iliev}:
\begin{equation}
res_z(\Delta^j\Phi(n;t',z)\Phi^*(n;t,z))=0, \qquad j\geq0.\end{equation}
Furthermore, we can rewrite the tau function to the following identity as follows\cite{LiuS,LiuSY,MPV}:\\
\begin{equation}
\oint(\Phi(m;t',z)\Phi^*(n;t,z))\frac{dz}{2\pi i}=0, \qquad m\geq n.
\end{equation}

After the relevant knowledge of discrete KP hierarchy, the addition formula of the discrete KP hierarchy can be declared. Set:

$[\alpha]=(\alpha,\frac{\alpha^2}{2},\frac{\alpha^3}{3},\cdots),$\quad \ \ \ \ \ $\xi(t,z)=\sum^\infty_{n=1}t_n z^n,$\quad \ \ \ \ \   $t=(t_1,t_2,\cdots).$\\
The $\tau(n;t)$ function of the discrete KP hierarchy is given by (\ref{f217}) and (\ref{f218}). Then we have
\begin{equation}\label{f221}
\oint e^{\xi(t'-t,z)}\tau(m,t'-[z^{-1}])\tau(n,t+[z^{-1}]) (1+z)^{m-n}\frac{dz}{2\pi i}=0,
\end{equation}\\
where $\oint$ is a formal algebraic operator, which collects the coefficient of $z^{-1}$ of Laurent series:
\begin{equation}
\oint \frac{dz}{2\pi i} \sum^\infty_{-\infty}a_nz^n=a_{-1}.\end{equation}\\
Let $t=x+y$, $t'=x-y$\cite{Yoko}. Then eq.(\ref{f221}) becomes
\begin{equation}\label{f223}
\oint e^{-2\xi(y,z)}\tau(m,x-y-[z^{-1}])\tau(n,x+y+[z^{-1}]) (1+z)^{m-n}\frac{dz}{2\pi i}=0.
\end{equation}
Suppose
\begin{equation}
y=\frac{1}{2}(\sum^{k-1}_{i=1}[\beta_i]-\sum^{k+1}_{i=1}[\alpha_i])\end{equation}
 in eq.(\ref{f223}), where $k\geq2, \quad k\in\mathbb{Z}.$ Then eq.(\ref{f223}) becomes
\begin{align}\label{f225}
\oint \exp(-\xi(\sum^{k-1}_{i=1}[\beta_i]-\sum^{k+1}_{i=1}[\alpha_i],z)) \tau(m,x-\frac{1}{2}(\sum^{k-1}_{i=1}[\beta_i]-\sum^{k+1}_{i=1}[\alpha_i])-[z^{-1}]) \\ \nonumber
\times  \tau(n,x+\frac{1}{2}(\sum^{k-1}_{i=1}[\beta_i]-\sum^{k+1}_{i=1}[\alpha_i])+[z^{-1}]) (1+z)^{m-n}\frac{dz}{2\pi i}=0.
\end{align}\\
Firstly we collect the exponential factor in eq.(\ref{f225}),
\begin{equation}
\exp(-\xi(\sum^{k-1}_{i=1}[\beta_i]-\sum^{k+1}_{i=1}[\alpha_i],z))=\frac{\prod^{k-1}_{i=1}(1-\beta_iz)}
{\prod^{k+1}_{i=1}(1-\alpha_iz)}.\end{equation}\\
Taking residues at $z=\alpha_i^{-1}$ $(1\leq i\leq {k+1})$ in the integral\cite{Sato}, we shift the variable $x$ as
$
x \rightarrow x+\frac{1}{2}(\sum^{k-1}_{i=1}[\beta_i]+\sum^{k+1}_{i=1}[\alpha_i])$,
we get the follow theorem.\\
{\sl {\bf  Theorem 1}} The addition formula for the $\tau$-function of the discrete KP hierarchy is:
\begin{equation}\label{f227}
\sum^{k+1}_{i=1}(1+\alpha_i^{-1})^{m-n}(-1)^{i-1}\zeta(n;x,\beta_1,\beta_2,\cdots,\beta_{k-1},\alpha_i)\zeta(m;x,\alpha_1,\alpha_2,\cdots,\hat \alpha_i,\cdots,\alpha_{k+1})=0,\end{equation}
where
\begin{equation}
\zeta(*;x,\alpha_1,\cdots,\alpha_n)=\hbar(\alpha_1,\cdots,\alpha_n)\tau(*;x+[\alpha_1]+\cdots+[\alpha_n]),\end{equation}
\begin{equation}
\hbar(\alpha_1,\cdots,\alpha_n)=\prod_{i<j}(\alpha_i-\alpha_j),\end{equation}
which $\hat \alpha_i$ means that $\alpha_i$ must be removed.

In order to show the addition formula of the discrete KP hierarchy clearly, we give the addition formula in the case of $k=2$ as the following examples:\\
{\sl {\bf  Example 1}} The tau function of discrete KP hierarchy satisfies the following identity in the case of $k=2$ and $m=n$:
\begin{align}\label{f230}
 \alpha_{01}\alpha_{23}\tau(n;x+[\alpha_0]+[\alpha_1])\tau(n;x+[\alpha_2]+[\alpha_3])-
 \alpha_{02}\alpha_{13}\tau(n;x+[\alpha_0]+[\alpha_2])\\ \nonumber\times \tau(n;x+[\alpha_1]+[\alpha_3])
+\alpha_{03}\alpha_{12}\tau(n;x+[\alpha_0]+[\alpha_3])\tau(n;x+[\alpha_1]+[\alpha_2])=0,\end{align}
where $\alpha_{ij}=\alpha_i-\alpha_j.$ It is the fact that the identity eq.(\ref{f230}) is the Fay identity of the classical KP hierarchy\cite{Dickey} in this case.\\
{\sl {\bf  Example 2}} The tau function of discrete KP hierarchy satisfies the following identity in the case of $k=2$ and $m\neq n$:
\begin{align}\label{f231}
 &\alpha_{01}\alpha_{23}(1+\alpha_1^{-1})^{m-n}\tau(n;x+[\alpha_0]+[\alpha_1])\tau(m;x+[\alpha_2]+[\alpha_3])\\ \nonumber
 &-\alpha_{02}\alpha_{13}(1+\alpha_2^{-1})^{m-n}\tau(n;x+[\alpha_0]+[\alpha_2])\tau(m;x+[\alpha_1]+[\alpha_3])\\ \nonumber
&+\alpha_{03}\alpha_{12}(1+\alpha_3^{-1})^{m-n}\tau(n;x+[\alpha_0]+[\alpha_3])\tau(m;x+[\alpha_1]+[\alpha_2])=0.\end{align}

Then we get the fact that eq.(\ref{f231}) is the Fay identity of the discrete KP hierarchy. To show the fact that the discrete KP hierarchy eq.(\ref{f221}) is equivalent to the identity eq.(\ref{f231}), we need the following three lemmas.\\
{\sl {\bf  Lemma 2.1}}  The following formula follows from eq.(\ref{f231}):
\begin{align}\label{f232}
&\frac{(1+\beta_1^{-1})^{m-n}\tau(n;x+\sum^{k}_{i=1}[\beta_i]-\sum^{k}_{i=1}[\alpha_i])}{\tau(m;x)}=\frac{\prod^k_{i,j=1}(\beta_i-\alpha_j)}{\prod_{i<j}\alpha_{ij}\beta_{ji}} \\ \nonumber \times det &\left (\begin{matrix} \frac{(1+\alpha_1^{-1})^{m-n}\tau(m;x+[\beta_1]-[\alpha_1])}{(\beta_1-\alpha_1)\tau(m;x)} &\frac{(1+\alpha_2^{-1})^{m-n}\tau(m;x+[\beta_1]-[\alpha_2])}{(\beta_1-\alpha_2)\tau(m;x)} & \dots &\frac{(1+\alpha_k^{-1})^{m-n}\tau(m;x+[\beta_1]-[\alpha_k])}{(\beta_1-\alpha_k)\tau(m;x)}\\
\frac{\tau(n;x+[\beta_2]-[\alpha_1])}{(\beta_2-\alpha_1)\tau(m;x)} & \frac{\tau(n;x+[\beta_2]-[\alpha_2])}{(\beta_2-\alpha_2)\tau(m;x)}  & \dots & \frac{\tau(n;x+[\beta_2]-[\alpha_k])}{(\beta_2-\alpha_k)\tau(m;x)}\\
\vdots & \vdots & \ddots & \vdots\\
\frac{\tau(n;x+[\beta_k]-[\alpha_1])}{(\beta_k-\alpha_1)\tau(m;x)} &\frac{\tau(n;x+[\beta_k]-[\alpha_2])}{(\beta_k-\alpha_2)\tau(m;x)} & \dots & \frac{\tau(n;x+[\beta_k]-[\alpha_k])}{(\beta_k-\alpha_k)\tau(m;x)} \end{matrix}\right),\end{align}
which $k\geq2$.

\begin{proof}
We change the names of $\alpha$ variables in eq.(\ref{f231}) as $(\alpha_3,\alpha_0) \rightarrow (\beta_1,\beta_2)$ and shift the variable $x$ as
$x \rightarrow x-[\alpha_1]-[\alpha_2]$. Then we solve it in $\tau(n;x+[\beta_1]+[\beta_2]-[\alpha_1]-[\alpha_2])$. We will get the case $k=2$ in the eq.(\ref{f232}).

Next use the mathematical induction to prove the above conclusion. First suppose that eq.(\ref{f232}) holds in the case of $k=s$:
\begin{align}\label{f233}
&(1+\beta_1^{-1})^{m-n} \tau(n;x+\sum^{s}_{i=1}[\beta_i]-\sum^{s}_{i=1}[\alpha_i])={\tau(m;x)}^{-s+1}C_s
\\ \nonumber &\times det \left (\begin{matrix} \frac{(1+\alpha_1^{-1})^{m-n}\tau(m;x+[\beta_1]-[\alpha_1])}{(\beta_1-\alpha_1)} &\frac{(1+\alpha_2^{-1})^{m-n}\tau(m;x+[\beta_1]-[\alpha_2])}{(\beta_1-\alpha_2)} & \dots &\frac{(1+\alpha_s^{-1})^{m-n}\tau(m;x+[\beta_1]-[\alpha_s])}{(\beta_1-\alpha_s)}\\
\frac{\tau(n;x+[\beta_2]-[\alpha_1])}{(\beta_2-\alpha_1)} & \frac{\tau(n;x+[\beta_2]-[\alpha_2])}{(\beta_2-\alpha_2)}  & \dots & \frac{\tau(n;x+[\beta_2]-[\alpha_s])}{(\beta_2-\alpha_s)}\\
\vdots & \vdots & \ddots & \vdots\\
\frac{\tau(n;x+[\beta_s]-[\alpha_1])}{(\beta_s-\alpha_1)} &\frac{\tau(n;x+[\beta_s]-[\alpha_2])}{(\beta_s-\alpha_2)} & \dots & \frac{\tau(n;x+[\beta_s]-[\alpha_s])}{(\beta_s-\alpha_s)} \end{matrix}\right), \end{align}
where
\begin{equation}
C_s=\frac{\prod^s_{i,j=1}(\beta_i-\alpha_j)}{\prod_{i<j}\alpha_{ij}\beta_{ji}}.\end{equation}
Shift the variable $x$ in eq.(\ref{f233}) as
\begin{equation}
x \rightarrow x+[\beta_{s+1}]-[\alpha_{s+1}].\end{equation}
Then the following equation can be derived
\begin{align}\label{f236}
&(1+\beta_1^{-1})^{m-n} \tau(n;x+\sum^{s+1}_{i=1}[\beta_i]-\sum^{s+1}_{i=1}[\alpha_i])={\tau(m;x+[\beta_{s+1}]-[\alpha_{s+1}])}^{-s+1}C_s \\ \nonumber
&\times det \left (\begin{matrix} \frac{(1+\alpha_1^{-1})^{m-n}\tau(m;x+[\beta_{s+1}]-[\alpha_{s+1}]+[\beta_1]-[\alpha_1])}{(\beta_1-\alpha_1)} & \dots &\frac{(1+\alpha_s^{-1})^{m-n}\tau(m;x+[\beta_{s+1}]-[\alpha_{s+1}]+[\beta_1]-[\alpha_s])}{(\beta_1-\alpha_s)}\\
\vdots & \ddots & \vdots\\
\frac{\tau(n;x+[\beta_{s+1}]-[\alpha_{s+1}]+[\beta_s]-[\alpha_1])}{(\beta_s-\alpha_1)} & \dots & \frac{\tau(n;x+[\beta_{s+1}]-[\alpha_{s+1}]+[\beta_s]-[\alpha_s])}{(\beta_s-\alpha_s)} \end{matrix}\right).\end{align}\\
With the above conclusion holding in the case of $k=2$, firstly, we observe the first row of the matrix in eq.(\ref{f236}),
\begin{equation}\label{f237}
\tau(m;x+[\beta_1]+[\beta_{s+1}]-[\alpha_j]-[\alpha_{s+1}])=\frac{\tau(m;x)^{-1}}{(1+\beta_1^{-1})^{m-n}}A_{1j}B_{1j},\end{equation}\\
where
\begin{equation}
A_{1j}=\frac{(\beta_1-\alpha_j)(\beta_1-\alpha_{s+1})(\beta_{s+1}-\alpha_j)(\beta_{s+1}-\alpha_{s+1})}{(\alpha_j-\alpha_{s+1})(\beta_{s+1}-\beta_1)},\end{equation}\\
\begin{equation}
B_{1j}=det \left(\begin{matrix} \frac{(1+\alpha_j^{-1})^{m-n} \tau(m;x+[\beta_1]-[\alpha_j])}{(\beta_1-\alpha_j)} & \frac{(1+\alpha_{s+1}^{-1})^{m-n} \tau(m;x+[\beta_1]-[\alpha_{s+1}])}{(\beta_1-\alpha_{s+1})}\\
\frac{\tau(m;x+[\beta_{s+1}]-[\alpha_j])}{(\beta_{s+1}-\alpha_j)} &\frac{\tau(m;x+[\beta_{s+1}]-[\alpha_{s+1}])}{(\beta_{s+1}-\alpha_{s+1})} \end{matrix}\right).\end{equation}
Next, we observe the other rows of the matrix in eq.(\ref{f236}),
\begin{equation}\label{f240}
\tau(n;x+[\beta_i]+[\beta_{s+1}]-[\alpha_j]-[\alpha_{s+1}])=\frac{\tau(m;x)^{-1}}{(1+\beta_1^{-1})^{m-n}}A_{ij}B_{ij},\end{equation}\\
where
\begin{equation}
A_{ij}=\frac{(\beta_i-\alpha_j)(\beta_i-\alpha_{s+1})(\beta_{s+1}-\alpha_j)(\beta_{s+1}-\alpha_{s+1})}{(\alpha_j-\alpha_{s+1})(\beta_{s+1}-\beta_i)},\end{equation}\\
\begin{equation}
B_{ij}=det \left(\begin{matrix} \frac{(1+\alpha_j^{-1})^{m-n} \tau(m;x+[\beta_i]-[\alpha_j])}{(\beta_i-\alpha_j)} & \frac{(1+\alpha_{s+1}^{-1})^{m-n} \tau(m;x+[\beta_i]-[\alpha_{s+1}])}{(\beta_i-\alpha_{s+1})}\\
\frac{\tau(n;x+[\beta_{s+1}]-[\alpha_j])}{(\beta_{s+1}-\alpha_j)} &\frac{\tau(n;x+[\beta_{s+1}]-[\alpha_{s+1}])}{(\beta_{s+1}-\alpha_{s+1})} \end{matrix}\right).\end{equation}\\
By substituting eq.(\ref{f237}) and eq.(\ref{f240}) into the determinant in the right hand side of eq.(\ref{f236}), we get
\begin{align}\label{f243}
&det \left (\begin{matrix} \frac{(1+\alpha_1^{-1})^{m-n}\tau(m;x+[\beta_{s+1}]-[\alpha_{s+1}]+[\beta_1]-[\alpha_1])}{(\beta_1-\alpha_1)} & \dots &\frac{(1+\alpha_s^{-1})^{m-n}\tau(m;x+[\beta_{s+1}]-[\alpha_{s+1}]+[\beta_1]-[\alpha_s])}{(\beta_1-\alpha_s)}\\
\vdots & \ddots & \vdots\\
\frac{\tau(n;x+[\beta_{s+1}]-[\alpha_{s+1}]+[\beta_s]-[\alpha_1])}{(\beta_s-\alpha_1)} & \dots & \frac{\tau(n;x+[\beta_{s+1}]-[\alpha_{s+1}]+[\beta_s]-[\alpha_s])}{(\beta_s-\alpha_s)} \end{matrix}\right)  \\ \nonumber
&=(\frac{\beta_{s+1}-\alpha_{s+1}}{\tau(m;x)})^s\prod^s_{i=1}\frac{(\beta_{s+1}-\alpha_i)(\beta_i-\alpha_{s+1})}{(\alpha_i-\alpha_{s+1})(\beta_{s+1}-\beta_i)}det(B_{ij})_{1\leq{i,j}\leq s}.\end{align}
By the Sylvester's theorem for determinants\cite{Yoko}, we get
\begin{align}\label{f244}
&det(B_{ij})_{1\leq{i,j}\leq s}=(\frac{\tau(m;x+[\beta_{s+1}]-[\alpha_{s+1}])}{\beta_{s+1}-\alpha_{s+1}})^{s-1}\\ \nonumber &\times det \left (\begin{matrix} \frac{(1+\alpha_1^{-1})^{m-n}\tau(m;x+[\beta_1]-[\alpha_1])}{(\beta_1-\alpha_1)} &\frac{(1+\alpha_2^{-1})^{m-n}\tau(m;x+[\beta_1]-[\alpha_2])}{(\beta_1-\alpha_2)} & \dots &\frac{(1+\alpha_{s+1}^{-1})^{m-n}\tau(m;x+[\beta_1]-[\alpha_{s+1}])}{(\beta_1-\alpha_{s+1})}\\
\frac{\tau(n;x+[\beta_2]-[\alpha_1])}{(\beta_2-\alpha_1)} & \frac{\tau(n;x+[\beta_2]-[\alpha_2])}{(\beta_2-\alpha_2)}  & \dots & \frac{\tau(n;x+[\beta_2]-[\alpha_{s+1}])}{(\beta_2-\alpha_{s+1})}\\
\vdots & \vdots & \ddots & \vdots\\
\frac{\tau(n;x+[\beta_{s+1}]-[\alpha_1])}{(\beta_{s+1}-\alpha_1)} &\frac{\tau(n;x+[\beta_{s+1}]-[\alpha_2])}{(\beta_{s+1}-\alpha_2)} & \dots & \frac{\tau(n;x+[\beta_{s+1}]-[\alpha_{s+1}])}{(\beta_{s+1}-\alpha_{s+1})} \end{matrix}\right).\end{align}\\
By substituting eq.(\ref{f243}) and eq.(\ref{f244}) into eq.(\ref{f236}), we drive the case of $k=s+1$ of eq.$(2.32)$. Then eq.(\ref{f232}) holds for any (\ref{f232}).\end{proof}
The difference of the addition formula between the KP hierarchy and the discrete KP hierarchy is the factor $(1+\alpha_i^{-1})^{m-n}$, such as the case $m=n$(\ref{f230}) and $m\neq n$(\ref{f231}). Because of the factor, it is difficult to prove the equation. Therefore, we divide the right hand side of eq.(\ref{f236}) into two section: the first row and the other rows. We use the case of $k=2$ in each section. Next we use the Pl\"ucker's relations for the determinant to simplify the determinant of the right hand side of eq.(\ref{f243})\cite{Yoko}.\\
{\sl {\bf  Lemma 2.2}} The addition formula (\ref{f227}) can be calculated from the Pl\"ucker's relations for the determinant of the right hand side of eq.(\ref{f232}).
\begin{proof}
Considering $k$ is fixed and the $\infty\times k$ matrix $T=(a_{ij})$ is
\begin{equation}
a_{1j}=\frac{(1+\beta_1^{-1})^{m-n} \tau(m;x+[\beta_1]-[\alpha_j])}{(\beta_1-\alpha_j)\tau(m;x)}, \quad  for \quad i=1, \end{equation}  and
\begin{equation}
a_{ij}=\frac{\tau(n;x+[\beta_i]-[\alpha_j])}{(\beta_i-\alpha_j)\tau(m;x)}, \quad  for \quad 2\leq i \leq k.\end{equation} \\
Then $\Gamma(p_1,\cdots, p_{k-1},l_i)$ and $\Gamma(l_1,\cdots,\hat{l_i},\cdots,, l_{k+1})$ can be expressed by $2k$ point functions by eq.(\ref{f232}). Then shift them to $(17)$ in the \cite{Yoko} and substitute the variable $x$ as
\begin{equation}
x\rightarrow x+\sum^k_{r=1}[\alpha_r],\end{equation}
we have the addition formula (\ref{f227}) by renaming the variables as $(\beta_{p_1},\cdots, \beta_{p_{k-1}})\rightarrow(\beta_1,\cdots, \beta_{k-1})$, $(\beta_{l_1},\cdots, \beta_{l_{k+1}})\rightarrow(\alpha_1,\cdots, \alpha_{k+1})$.
 \end{proof}

Because the eq.(\ref{f221}) is equivalent to eq.(\ref{f223}), by setting the left hand side of eq.(\ref{f223}) to be $F(y)$, the next lemma can be proved similarly as \cite{Yoko}.\\
{\sl {\bf  Lemma 2.3}} The discrete KP hierarchy eq.(\ref{f221}) is equivalent to (\ref{f227}).

Owing to lemma $(2.1)$, $(2.2)$ and $(2.3)$, we get the fact that the discrete KP hierarchy eq.(\ref{f221}) is equivalent to the identity eq.(\ref{f231}). Above we study the addition formula for the $\tau$-function of the discrete KP hierarchy, next we are going to learn another integrable system, namely q-deformed KP hierarchy.

%\begin{center}
\section{The addition formula of the q-KP hierarchy }
%\end{center}
\setcounter{section}{3}

The q-deformed KP (q-KP) hierarchy is also an important object in the research of the integrable systems\cite{DKJM,Sato,Dickey,Segal}. The q-deformed integrable system is a system about the q-derivative $D_q$ instead of the usual derivative with respect to $x$ in the classical system. Our purpose in this paper is to prove the addition formula of the q-KP hierarchy. We should recall some basic known facts about the q-KP hierarchy\cite{Jingsong,PLAMEN,Kelei}. Firstly we introduce the Euler-Jackson q-derivative $D_q$ which is
\begin{equation}
D_qf(x)=\frac{f(xq)-f(x)}{x(q-1)},\end{equation}
and the q-shift operator is
\begin{equation}
Df(x)=f(xq).\end{equation}
Obviously, $D^k$ for any $k\in\mathbb{Z}$ is $D^kf=f(xq^k)$. We have the $D_q^{-1}$ as the formal inverse of $D_q$, then the following q-deformed Leibnitz rule for any $n\in\mathbb{Z}$ holds true as the q-pseudo-difference operator
\begin{equation}
D_q^n\circ f=\sum^{\infty}_{k=0}\binom{n}{k}_qD^{n-k}(D_q^kf)D_q^{n-k},
\end{equation}
where the q-binomial coefficients for $n\in\mathbb{Z}$, $k\in\mathbb{Z}_+$ are denoted as
\begin{equation}
\binom{n}{0}_q=1, \qquad \binom{n}{k}_q=\frac{(n)_q(n-1)_q\cdots(n-k+1)_q}{(1)_q(2)_q\cdots(k)_q},
\end{equation}
which we set the q-number as $(n)_q=\frac{q^n-1}{q-1}$.
The adjoint operator $D_q^*$ can be defined as:
\begin{equation}
D_q^*=-D_qD^{-1}=-\frac{1}{q}D_{\frac{1}{q}}.\end{equation}

Similar to KP hierarchy, we have the following properties of the wave function and tau function.
The q-KP hierarchy\cite{PLAMEN} is a family of evolution equations depending on infinitely many variables $t=(t_1,t_2,\cdots)$:
\begin{equation}
\frac{\partial L}{\partial t_n}=[B_n, L],\ \ \ B_n:=(L^n)_+,
\end{equation}
where $L$ is the formal q-pseudo-difference operator
\begin{equation} \label{laxoperatordkp}
L=D_q + u_0+u_{-1}D_q^{-1}+u_{-2}D_q^{-2}+\cdots,
\end{equation}
with $u_i=u_i(x,t_1,t_2,\cdots), i=0,-1,-2,-3,\cdots$.\\
Define the following dressing operator:
\begin{equation}
S=1+\sum_{j=1}^{\infty}s_jD_q^{-j}\end{equation}
which satisfies
\begin{equation}
L=SD_qS^{-1}.\end{equation}
The dressing operator $S$ satisfies the Sato equation
\begin{equation}
\frac{\partial S}{\partial t_n}=-(L^n)_-S, \qquad  n=1,2,3,\cdots.\end{equation}
We set $(n)_q!=(n)_q(n-1)_q(n-2)_q\cdots (1)_q$ and quote the q-exponent $e_q(x)$ as follows
\begin{equation}
e_q(x)=\sum_{n=0}^{\infty}\frac{x^n}{(n)_q!}=\exp(\sum_{k=0}^{\infty}\frac{(1-q)^k}{k(1-q^k)}x^k),\end{equation}
and we introduce the notation
\begin{equation}
P|_{\frac{x}{t}}=\sum_ip_i(\frac{x}{t})t^iD_q^i,\end{equation}
for a q-pseudo-difference operator $P=\sum_ip_iD_q^i$.
Then we quote the q-wave function $w_q(x;t,z)$ and the adjoint q-wave
function $w_q^*(x;t,z)$:
\[\label{w}
w_q(x;t,z)=S e_q(xz)\exp(\sum^\infty_{i=1}t_i z^i),
\]
and
\[\label{w}
w_q^*(x;t,z)=(S^*)^{-1}|_{x/q} e_{1/q}(-xz) \exp(\sum^\infty_{i=1}-t_i
z^i),
\]
which satisfies
\[
Lw_q=zw_q,\ \ \ L^*|_{x/q}w_q^*=zw_q^*,
\]
and
\[
\d_{t_i}w_q=B_iw_q,\ \ \ \d_{t_i}w_q^*=-B_i^*|_{x/q}w_q^*.
\]
Also we use a tau function $\tau_q(x;t)$ for the q-KP hierarchy \cite{Jingsong}, which satisfies
\begin{equation}\label{taudefinition1}
w_q(x;t,z)=\frac{\tau_q(x;t-[z^{-1}])}{\tau_q(x;t)}e_q(xz)
\exp(\sum^\infty_{i=1}t_i z^i),
\end{equation}
and
\begin{equation}\label{taudefinition2}
w_q^*(x;t,z)=\frac{\tau_q(x;t+[z^{-1}])}{\tau_q(x;t)}e_{1/q}(-xz)
\exp(\sum^\infty_{i=1}-t_i z^i),
\end{equation}
where $[z]=(z,\frac{z^2}{2},\frac{z^3}{3},\cdots)$.

From the above conclusion, we get the following bilinear identities. The wave function and adjoint wave function satisfy the following bilinear identities.
Hirota bilinear identities of the q-KP hierarchy for $n\in\mathbb{Z}_+$ and any $\alpha=(\alpha_1, \alpha_2, \cdots)$ are\cite{PLAMEN}
\begin{equation}
\oint D_q^n\partial^{\alpha}w_q(x;t',z)w_q^*(x;t,z)\frac{dz}{2\pi i}=0,\end{equation}
where $\partial^{\alpha}=\partial_{t_1}^{\alpha_1}\partial_{t_2}^{\alpha_2}\cdots \partial_{t_k}^{\alpha_k}.$
Furthermore, we change the tau function to the following equations as follows\cite{LiuS,MHTu}:
\begin{equation}
\oint(\tilde w_q(x;t',z)\tilde w_q^*(x;t,z))\frac{dz}{2\pi i}=0.
\end{equation}\\
$\tilde w_q(x;t',z)$ and $\tilde w_q^*(x;t,z)$ are also wave functions of the q-KP hierarchy, so let
\begin{equation}\label{f321}
\tilde w_q(x;t',z)=\frac{\tau_q(x;t'-[z^{-1}])}{\tau_q(x;t')}e_q^{xz}
\exp(\sum^\infty_{i=1}t'_i z^i),
\end{equation}
and
\begin{equation}\label{f322}
\tilde w_q^*(x;t,z)=\frac{\tau_q(x;t+[z^{-1}])}{ \tau_q(x;t)}(e_q^{xz})^{-1}
\exp(\sum^\infty_{i=1}-t_i z^i),
\end{equation}
where $t-[z^{-1}]=(t_1-\frac{1}{z},t_2-\frac{1}{2z^2},t_3-\frac{1}{3z^3},\cdots)$. Above we introduce the relevant knowledge of the q-KP hierarchy. Next, we will deduce its addition formula concretely. Set:

\quad $[\alpha]=(\alpha,\frac{\alpha^2}{2},\frac{\alpha^3}{3},\cdots),$\quad \ \ \ \ \ $\xi(t,z)=\sum^\infty_{n=1}t_n z^n,$\quad \ \ \ \ \   $t=(t_1,t_2,\cdots).$\\
The $\tau_q(x;t)$ function of the q-KP hierarchy is given by (\ref{f321}) and (\ref{f322}). Then we have
\begin{equation}\label{f323}
\oint e^{\xi(t'-t,z)} \tau_q(x;t'-[z^{-1}]) \tau_q(x;t+[z^{-1}])\frac{dz}{2\pi i}=0.
\end{equation}
Let $t=u+v$, $t'=u-v$, then eq.(\ref{f323}) becomes
\begin{equation}\label{f324}
\oint e^{-2\xi(v,z)} \tau_q(x;u-v-[z^{-1}]) \tau_q(x;u+v+[z^{-1}])\frac{dz}{2\pi i}=0.
\end{equation}
Suppose
\begin{equation}
v=\frac{1}{2}(\sum^{k-1}_{i=1}[\beta_i]-\sum^{k+1}_{i=1}[\alpha_i])\end{equation}
 in eq.(\ref{f324}), where $k\geq2$, $k\in\mathbb{Z}$. Then eq.(\ref{f324}) becomes
\begin{align}\label{f326}
\oint \exp(-\xi(\sum^{k-1}_{i=1}[\beta_i]-\sum^{k+1}_{i=1}[\alpha_i],z))  \tau_q(x;u-\frac{1}{2}(\sum^{k-1}_{i=1}[\beta_i]-\sum^{k+1}_{i=1}[\alpha_i])-[z^{-1}]) \\ \nonumber
\times \tau_q(x;u+\frac{1}{2}(\sum^{k-1}_{i=1}[\beta_i]-\sum^{k+1}_{i=1}[\alpha_i])+[z^{-1}])\frac{dz}{2\pi i}=0.
\end{align}
Firstly we collect the exponential factor in eq.(\ref{f326}) as in the discrete KP hierarchy,
\begin{equation}
\exp(-\xi(\sum^{k-1}_{i=1}[\beta_i]-\sum^{k+1}_{i=1}[\alpha_i],z))=\frac{\prod^{k-1}_{i=1}(1-\beta_iz)}
{\prod^{k+1}_{i=1}(1-\alpha_iz)}.\end{equation}\\
Taking residues at $z=\alpha_i^{-1}$$(1\leq i\leq {k+1})$ in the integral, we shift the variable $u$ as $u \rightarrow u+\frac{1}{2}(\sum^{k-1}_{i=1}[\beta_i]+\sum^{k+1}_{i=1}[\alpha_i])$, we get the following theorem.\\
{\sl {\bf  Theorem 2}} The addition formula for the $\tau$-function of the q-KP hierarchy is:
\begin{equation}\label{f328}
\sum^{k+1}_{i=1}(-1)^{i-1}\zeta_q(x,\beta_1,\beta_2,\cdots,\beta_{k-1},\alpha_i)\zeta_q(x,\alpha_1,\alpha_2,\cdots,\hat \alpha_i,\cdots,\alpha_{k+1})=0,\end{equation}\\
where
\begin{equation}
\zeta_q(x,\alpha_1,\cdots,\alpha_n)=\hbar(\alpha_1,\cdots,\alpha_n)\tau_q(x;u+[\alpha_1]+\cdots+[\alpha_n]),\end{equation}
\begin{equation}
\hbar(\alpha_1,\cdots,\alpha_n)=\prod_{i<j}(\alpha_i-\alpha_j).\end{equation}

In order to show the addition formula of the q-KP hierarchy clearly, we give the addition formula in the case of $k=2$ as the following example:\\\
{\sl {\bf  Example 3}} The tau function of q-KP hierarchy satisfies the following identity in the case of $k=2$:
\begin{align}\label{f331}
 \alpha_{01}\alpha_{23}\tau_q(x;u+[\alpha_0]+[\alpha_1])\tau_q(x;u+[\alpha_2]+[\alpha_3])-
 \alpha_{02}\alpha_{13}\tau_q(x;u+[\alpha_0]+[\alpha_2])\\ \nonumber\times \tau_q(x;u+[\alpha_1]+[\alpha_3])
+\alpha_{03}\alpha_{12}\tau_q(x;u+[\alpha_0]+[\alpha_3])\tau_q(x;u+[\alpha_1]+[\alpha_2])=0.\end{align}

Then we obtain the fact that eq.(\ref{f331}) is the Fay identity of the q-KP hierarchy, we also show that the q-KP hierarchy eq.(\ref{f323}) is equivalent to the identity eq. (\ref{f331}) by the following lemmas as the similar way of the discrete KP hierarchy case.\\
{\sl {\bf  Lemma 3.1}}   The following formula follows from eq.(\ref{f331}):
\begin{align}\label{f332}
&\frac{\tau_q(x;u+\sum^{k}_{i=1}[\beta_i]-\sum^{k}_{i=1}[\alpha_i])}{\tau_q(x;u)}\\ \nonumber &=\frac{\prod^k_{i,j=1}(\beta_i-\alpha_j)}{\prod_{i<j}\alpha_{ij}\beta_{ji}} det(\frac{\tau_q(x;u+[\beta_i]-[\alpha_j])}{(\beta_i-\alpha_j)\tau_q(x;u)})_{1\leq i,j\leq k}\end{align}\\
which $k\geq2$.\\
{\sl {\bf  Lemma 3.2}} The addition formula (\ref{f328}) can be calculated from the Pl\"ucker's relations for the determinant of the right hand side of eq.(\ref{f332}).\\
{\sl {\bf  Lemma 3.3}}   The q-KP hierarchy eq.(\ref{f323}) are equivalent to (\ref{f328}).

Owing to Lemma $(3.1)$, $(3.2)$ and $(3.3)$, we get the fact that the q-KP hierarchy eq.(\ref{f323}) is equivalent to the identity eq.(\ref{f331}).

In fact, the $\tau$-function $\tau_q$ of the q-KP hierarchy has some connection with the $\tau$-function $\tau$ of the classical KP hierarchy\cite{PLAMEN} as
\begin{equation}\label{f333}
\tau_q(x;t)=\tau(t+[x]_q),\end{equation}
where
\begin{equation}
[x]_q=(x,\frac{(1-q)^2}{2(1-q^2)}x^2,\frac{(1-q)^3}{3(1-q^3)}x^3,\cdots).\end{equation}
By the relationship (\ref{f333}) we easily get the lemmas $(3.1),(3.2),(3.3)$, and the addition formula $(3.28)$ from the addition formula of the KP hierarchy can be got\cite{Yoko}:
\begin{equation}
\sum^{m+1}_{i=1}(-1)^{i-1}\zeta(x;\beta_1,\beta_2,\cdots,\beta_{m-1},\alpha_i)\zeta(x;\alpha_1,\alpha_2,\cdots,\hat \alpha_i,\cdots,\alpha_{m+1})=0,\end{equation}\\
where
\begin{equation}
\zeta(x,\alpha_1,\cdots,\alpha_n)=\triangle(\alpha_1,\cdots,\alpha_n)\tau(x+[\alpha_1]+\cdots+[\alpha_n]),\end{equation}
\begin{equation}
\triangle(\alpha_1,\cdots,\alpha_n)=\prod_{i<j}(\alpha_i-\alpha_j).\end{equation}

Above studies reveal that the addition formula exists in the continuous integrable system, also exists in the discrete integrable system. Naturally, there is a problem that whether the addition formula will exist in the other KP type systems. The BKP hierarchy as the important sub-hierarchy is a hot topic in the research of the integrable systems. Next we will study the two-component BKP hierarchy and its reduction.

%\begin{center}
\section{The addition formula of the two-component BKP hierarchy }
%\end{center}
\setcounter{section}{4}

The two-component BKP hierarchy has two sets of time variables $t_1,t_3,\cdots, t_{2n+1}$ and $\bar{t}_1,\bar{t}_3,\cdots, \\ \bar{t}_{2n+1}$ with odd indices. The two-component BKP hierarchy is defined by two Lax operators. We set an algebra $\mathcal{A}$ as an algebra of smooth functions of a spatial coordinate $x$ and a derivation denoted as $\partial=\frac{d}{dx}$\cite{CZ Li,Chaozhong,SiQi}. One has the following multiplying rules:
\begin{equation}
\partial^i\cdot f=\sum_{r\geq 0}\binom{j}{i}\partial^r(f)\partial^{i-r},\qquad f\in \mathcal{A}.\end{equation}
Similar to the discrete KP hierarchy, the two-component BKP hierarchy can be defined by two Lax operators:
\begin{equation}
L=\partial+\sum_{i\geq 1}u_i\partial^{-i},\qquad \bar{L}=\partial^{-1}\bar{u}_{-1}+\sum_{i\geq 1}\bar{u}_i\partial^{i}.\end{equation}
The B type condition of two-component BKP hierarchy can be given by \cite{CZ Li}:
\begin{equation}
L^*=-\partial L\partial^{-1},\qquad \bar{L}^*=-\partial \bar{L}\partial^{-1},\qquad r\in \mathbb{Z}_+.\end{equation}
The Lax operators can be rewritten in a dressing form as
\begin{equation}
L=\Phi \partial\Phi^{-1},\qquad \bar{L}=\bar{\Phi}\partial^{-1}\bar{\Phi}^{-1},\end{equation}
which
\begin{equation}
\Phi=1+\sum_{i\geq 1}a_i\partial^{-i},\qquad \bar{\Phi}=1+\sum_{i\geq 1}b_i\partial^{i},\end{equation}
satisfy
\begin{equation}
\Phi^*=\partial\Phi^{-1}\partial^{-1},\qquad \bar{\Phi}^*=\partial\bar{\Phi}^{-1}\partial^{-1}.\end{equation}

The following equations can also be redefined by \cite{CZ Li,Chaozhong,SiQi}:
\begin{equation}\label{f47}
\frac{\partial\Phi}{\partial t_k}=-(L^k)_-\Phi,\qquad \frac{\partial\bar{\Phi}}{\partial t_k}=((L^k)_+-\delta_{k1}\bar{L}^{-1})\bar{\Phi},\end{equation}
\begin{equation}\label{f48}
\frac{\partial\Phi}{\partial \bar{t}_k}=-(\bar{L}^k)_-\Phi,\qquad \frac{\partial\bar{\Phi}}{\partial \bar{t}_k}=(\bar{L}^k)_+\bar{\Phi},\end{equation}
which $k\in \mathbb{Z}^{odd}_+$. Next one introduce two wave functions of $t=(t_1,t_3,\cdots, t_{2n+1},\cdots),\bar{t}=(\bar{t}_1,\bar{t}_3,\cdots, \bar{t}_{2n+1},\cdots)$
\begin{equation}
w(z)=w(t,\bar{t};z)=\Phi e^{\tilde{\xi}(t;z)},\qquad \bar{w}(\bar{z})=\bar{w}(t,\bar{t};\bar{z})=\bar{\Phi} e^{x\bar{z}+\tilde{\xi}(\bar{t};-\bar{z}^{-1})},\end{equation}
where $x=t_1$ and the function $\xi$ is defined as $\tilde{\xi}(t;z)=\sum_{k\in \mathbb{Z}^{odd}_+}t_kz^k$. The tau function of the two-component BKP hierarchy can be defined in form of the wave functions as
\begin{equation}
w(t,\bar{t};z)=\frac{\tau(t-2[z^{-1}],\bar{t})}{\tau(t,\bar{t})}e^{\tilde{\xi}(t;z)},\qquad \bar{w}(t,\bar{t};\bar{z})=\frac{\tau(t,\bar{t}-2[\bar{z}^{-1}])}{\tau(t,\bar{t})}e^{\tilde{\xi}(\bar{t};\bar{z})},\end{equation}
where $[z]=(z,\frac{z^3}{3},\frac{z^5}{5},\cdots)$.

From the above conclusion, we can obtain the following equation. The wave function and adjoint wave function satisfy the following bilinear identity.\\
The two-component BKP hierarchy is equivalent to the following bilinear equation\cite{Kanehisa}:
\begin{equation}\label{f411}
_zz^{-1}w(t',\bar{t'};z)w(t,\bar{t};-z)=res_{\bar{z}}\bar{z}^{-1}\bar{w}(t',\bar{t'};\bar{z})\bar{w}(t,\bar{t};-\bar{z}).\end{equation}
The $\tau$-function $\tau(t,\bar{t})$ of the two-component BKP hierarchy can be got by the bilinear equation\cite{Kanehisa}:
\begin{equation}\label{f412}
\oint\frac{dz}{2\pi iz}e^{\tilde{\xi}(t'-t,z)}\tau(t'-2[z^{-1}],\bar{t}')\tau(t+2[z^{-1}],\bar{t})=
\oint\frac{d\bar{z}}{2\pi i\bar{z}}e^{\tilde{\xi}(\bar{t}'-\bar{t},\bar{z})}\tau(t',\bar{t}'-2[\bar{z}^{-1}])\tau(t,\bar{t}+2[\bar{z}^{-1}]).\end{equation}\\

Above we introduce the relevant knowledge of the two-component BKP hierarchy. Next, we will deduce its addition formula concretely. Set:$[\alpha]_o=(\alpha,\frac{\alpha^3}{3},\frac{\alpha^5}{5},\cdots)$.

The $\tau(t,\bar{t})$ function of the two-component BKP hierarchy has been defined as eq.(\ref{f412}).
Then set $t=x+y$, $t'=x-y$, $\bar{t}=\bar{x}+\bar{y}$, $\bar{t}'=\bar{x}-\bar{y}$. Then eq.(\ref{f412}) becomes
\begin{align}\label{f413}
&\oint\frac{dz}{2\pi iz}e^{-2\tilde{\xi}(y,z)}\tau(x-y-2[z^{-1}],\bar{x}-\bar{y})\tau(x+y+2[z^{-1}],\bar{x}+\bar{y})\\ \nonumber =
&\oint\frac{d\bar{z}}{2\pi i\bar{z}}e^{-2\tilde{\xi}(\bar{y},\bar{z})}\tau(x-y,\bar{x}-\bar{y}-2[\bar{z}^{-1}])\tau(x+y,\bar{x}+\bar{y}+2[\bar{z}^{-1}]).\end{align}
Suppose
\begin{equation}
y=\sum^k_{i=1}[\alpha_i]_o,\qquad \bar{y}=\sum^k_{i=1}[\bar{\alpha_i}]_o,\end{equation}
in eq.(\ref{f413}). Then eq.(\ref{f413}) becomes
\begin{align}\label{f415}
\oint\frac{dz}{2\pi iz}e^{-2\tilde{\xi}(\sum^k_{i=1}[\alpha_i]_o,z)}\tau(x-\sum^k_{i=1}[\alpha_i]_o-2[z^{-1}],\bar{x}-\sum^k_{i=1}
[\bar{\alpha_i}]_o)\\ \nonumber \tau(x+\sum^k_{i=1}[\alpha_i]_o+2[z^{-1}],\bar{x}+\sum^k_{i=1}[\bar{\alpha_i}]_o)\\ \nonumber =
\oint\frac{d\bar{z}}{2\pi i\bar{z}}e^{-2\tilde{\xi}(\sum^k_{i=1}[\bar{\alpha_i}]_o,\bar{z})}\tau(x-\sum^k_{i=1}
[\alpha_i]_o,\bar{x}-\sum^k_{i=1}[\bar{\alpha_i}]_o-2[\bar{z}^{-1}])\\ \nonumber\tau(x+\sum^k_{i=1}[\alpha_i]_o,\bar{x}+\sum^k_{i=1}[\bar{\alpha_i}]_o+2[\bar{z}^{-1}]).\end{align}
Firstly sort out the exponential factor in eq.(\ref{f415}). By calculating $-2\sum^\infty_{n=1}t_{2n-1}\lambda^{2n-1}$ with the formula
\begin{equation}
-2\sum^\infty_{n=1}t_{2n-1}\lambda^{2n-1}=-\sum^\infty_{n=1}t_n\lambda^n+\sum^\infty_{n=1}t_n(-\lambda)^n,\end{equation}
we get
\begin{equation}\label{f417}
\exp(-2\tilde{\xi}(\sum^k_{i=1}[\alpha_i]_o,z))=\prod^k_{i=1}\frac{1-\alpha_iz}{1+\alpha_iz},\end{equation}
\begin{equation}\label{f418}
\exp(-2\tilde{\xi}(\sum^k_{i=1}[\bar{\alpha_i}]_o,\bar{z}))=\prod^k_{i=1}\frac{1-\bar{\alpha}_i\bar{z}}{1+\bar{\alpha}_i\bar{z}}.\end{equation}
Taking residues at $z=\alpha_i^{-1},\bar{z}=\bar{\alpha}_i^{-1}$$(1\leq i\leq {k})$ in the integral as before, we shift the variable $x$ as $x\rightarrow x+\sum^k_{i=1}[\alpha_i]_o,
\bar{x}\rightarrow \bar{x}+\sum^k_{i=1}[\bar{\alpha}_i]_o
$, we get the following theorem.\\
{\sl {\bf  Theorem 3}} The addition formula for the $\tau$-function of the two-component BKP hierarchy is:
\begin{align}\label{f419}
\sum^k_{i=1}\prod^k_{j=1,j \neq i}\frac{\alpha_i+\alpha_j}{\alpha_i-\alpha_j}\tau(x+2[\alpha_i]_o,\bar{x})
\tau(x+2\sum^k_{j=1,j \neq i}[\alpha_j]_o,\bar{x}+2\sum^k_{j=1}[\bar{\alpha}_j]_o) =\\ \nonumber
\sum^k_{i=1} \prod^k_{j=1,j \neq i}\frac{\bar{\alpha}_i+\bar{\alpha}_j}{\bar{\alpha}_i-\bar{\alpha}_j}\tau(x,\bar{x}+2[\bar{\alpha}_i]_o) \tau(x+2\sum^k_{j=1}[\alpha_j]_o,\bar{x}+2\sum^k_{j=1,j \neq i}[\bar{\alpha}_j]_o).\end{align}
The eq.(\ref{f419}) of the $\tau$-function will be simplified into the following equation by dividing $\tau(x,\bar{x})^2$.
The addition formula for the $\tau$-function of the two-component BKP hierarchy is:
\begin{align}\label{f420}
&\sum^k_{i=1}(-1)^{i-1}\frac{\tau(x+2[\alpha_i]_o,\bar{x})}{\tau(x,\bar{x})}A^{-1}_{1,\cdots,\hat{i},\cdots,k}\frac{\tau(x+2\sum^k_{j=1,j \neq i}[\alpha_j]_o,\bar{x}+2\sum^k_{j=1}[\bar{\alpha}_j]_o)}{\tau(x,\bar{x})}\\ \nonumber &=\frac{\bar{A}_{1,\cdots,k}}{A_{1,\cdots,k}}\sum^k_{i=1}(-1)^{i-1}\frac{\tau(x,\bar{x}+2[\bar{\alpha}_i]_o)}{\tau(x,\bar{x})}\bar{A}^{-1}_{1,\cdots,\hat{i},\cdots,k}\frac{\tau(x+2\sum^k_{j=1}[\alpha_j]_o,\bar{x}+2\sum^k_{j=1,j \neq i}[\bar{\alpha}_j]_o)}{\tau(x,\bar{x})},\end{align}
where  \\$A_{1,\cdots,k}=\prod^k_{i<j}\frac{\tilde{\alpha}_{ij}}{\alpha_{ij}},\qquad\bar{A}_{1,\cdots,k}=\prod^k_{i<j}\frac{\ddot{\alpha}_{ij}}{\dot{\alpha}_{ij}},$\\ $\tilde{\alpha}_{ij}=\alpha_i+\alpha_j,\qquad \alpha_{ij}=\alpha_i-\alpha_j,\qquad \ddot{\alpha}_{ij}=\bar{\alpha}_i+\bar{\alpha}_j,\qquad \dot{\alpha}_{ij}=\bar{\alpha}_i-\bar{\alpha}_j.$

In order to show the addition formula of the two-component BKP hierarchy clearly, we give the addition formula in the case of $k=3$ and $k=4$ as the following examples.\\
{\sl {\bf  Example 4}} The case $k=3$ of (\ref{f420}) becomes
\begin{align}
\frac{\tau(x+2[\alpha_1]_o,\bar{x})}{\tau(x,\bar{x})}\frac{\alpha_{23}}{\tilde{\alpha}_{23}}\frac{\tau(x+2[\alpha_2]_o+2[\alpha_3]_o,\bar{x}+2[\bar{\alpha}_1]_o+2[\bar{\alpha}_2]_o+2[\bar{\alpha}_3]_o)}{\tau(x,\bar{x})}\\ \nonumber
-\frac{\tau(x+2[\alpha_2]_o,\bar{x})}{\tau(x,\bar{x})}\frac{\alpha_{13}}{\tilde{\alpha}_{13}}\frac{\tau(x+2[\alpha_1]_o+2[\alpha_3]_o,\bar{x}+2[\bar{\alpha}_1]_o+2[\bar{\alpha}_2]_o+2[\bar{\alpha}_3]_o)}{\tau(x,\bar{x})}\\ \nonumber
+\frac{\tau(x+2[\alpha_3]_o,\bar{x})}{\tau(x,\bar{x})}\frac{\alpha_{12}}{\tilde{\alpha}_{12}}\frac{\tau(x+2[\alpha_1]_o+2[\alpha_2]_o,\bar{x}+2[\bar{\alpha}_1]_o+2[\bar{\alpha}_2]_o+2[\bar{\alpha}_3]_o)}{\tau(x,\bar{x})}\\ \nonumber
=\frac{\bar{A}_{123}}{A_{123}}\{\frac{\tau(x,\bar{x}+2[\bar{\alpha}_1]_o)}{\tau(x,\bar{x})}\frac{\dot{\alpha}_{23}}{\ddot{\alpha}_{23}}\frac{\tau(x+2[\alpha_1]_o+2[\alpha_2]_o+2[\alpha_3]_o,\bar{x}+2[\bar{\alpha}_2]_o+2[\bar{\alpha}_3]_o)}{\tau(x,\bar{x})}\\ \nonumber
-\frac{\tau(x,\bar{x}+2[\bar{\alpha}_2]_o)}{\tau(x,\bar{x})}\frac{\dot{\alpha}_{13}}{\ddot{\alpha}_{13}}\frac{\tau(x+2[\alpha_1]_o+2[\alpha_2]_o+2[\alpha_3]_o,\bar{x}+2[\bar{\alpha}_1]_o+2[\bar{\alpha}_3]_o)}{\tau(x,\bar{x})}\\ \nonumber
+\frac{\tau(x,\bar{x}+2[\bar{\alpha}_3]_o)}{\tau(x,\bar{x})}\frac{\dot{\alpha}_{12}}{\ddot{\alpha}_{12}}\frac{\tau(x+2[\alpha_1]_o+2[\alpha_2]_o+2[\alpha_3]_o,\bar{x}+2[\bar{\alpha}_1]_o+2[\bar{\alpha}_2]_o)}{\tau(x,\bar{x})}\}.\end{align}
{\sl {\bf  Example 5}} The case $k=4$ of (\ref{f420}) becomes
\begin{align}\label{f422}
&\frac{\tau(x+2[\alpha_1]_o,\bar{x})}{\tau(x,\bar{x})}\frac{\alpha_{23}}{\tilde{\alpha}_{23}}\frac{\alpha_{24}}{\tilde{\alpha}_{24}}\frac{\alpha_{34}}{\tilde{\alpha}_{34}}\frac{\tau(x+2[\alpha_2]_o+2[\alpha_3]_o+2[\alpha_4]_o,\bar{x}+2\sum^4_{j=1}[\bar{\alpha}_j]_o)}{\tau(x,\bar{x})}\\ \nonumber
&-\frac{\tau(x+2[\alpha_2]_o,\bar{x})}{\tau(x,\bar{x})}\frac{\alpha_{13}}{\tilde{\alpha}_{13}}\frac{\alpha_{14}}{\tilde{\alpha}_{14}}\frac{\alpha_{34}}{\tilde{\alpha}_{34}}\frac{\tau(x+2[\alpha_1]_o+2[\alpha_3]_o+2[\alpha_4]_o,\bar{x}+2\sum^4_{j=1}[\bar{\alpha}_j]_o)}{\tau(x,\bar{x})}\\ \nonumber
&+\frac{\tau(x+2[\alpha_3]_o,\bar{x})}{\tau(x,\bar{x})}\frac{\alpha_{12}}{\tilde{\alpha}_{12}}\frac{\alpha_{14}}{\tilde{\alpha}_{14}}\frac{\alpha_{24}}{\tilde{\alpha}_{24}}\frac{\tau(x+2[\alpha_1]_o+2[\alpha_2]_o+2[\alpha_4]_o,\bar{x}+2\sum^4_{j=1}[\bar{\alpha}_j]_o)}{\tau(x,\bar{x})}\\ \nonumber
&-\frac{\tau(x+2[\alpha_4]_o,\bar{x})}{\tau(x,\bar{x})}\frac{\alpha_{12}}{\tilde{\alpha}_{12}}\frac{\alpha_{13}}{\tilde{\alpha}_{13}}\frac{\alpha_{23}}{\tilde{\alpha}_{23}}\frac{\tau(x+2[\alpha_1]_o+2[\alpha_2]_o+2[\alpha_3]_o,\bar{x}+2\sum^4_{j=1}[\bar{\alpha}_j]_o)}{\tau(x,\bar{x})}\\ \nonumber
&=\frac{\bar{A}_{1234}}{A_{1234}}\\ \nonumber &\times \{\frac{\tau(x,\bar{x}+2[\bar{\alpha}_1]_o)}{\tau(x,\bar{x})}\frac{\dot{\alpha}_{23}}{\ddot{\alpha}_{23}}\frac{\dot{\alpha}_{24}}{\ddot{\alpha}_{24}}\frac{\dot{\alpha}_{34}}{\ddot{\alpha}_{34}}\frac{\tau(x+2\sum^4_{j=1}[\alpha_j]_o,\bar{x}+2[\bar{\alpha}_2]_o+2[\bar{\alpha}_3]_o+2[\bar{\alpha}_4]_o)}{\tau(x,\bar{x})}\\ \nonumber
&-\frac{\tau(x,\bar{x}+2[\bar{\alpha}_2]_o)}{\tau(x,\bar{x})}\frac{\dot{\alpha}_{13}}{\ddot{\alpha}_{13}}\frac{\dot{\alpha}_{14}}{\ddot{\alpha}_{14}}\frac{\dot{\alpha}_{34}}{\ddot{\alpha}_{34}}\frac{\tau(x+2\sum^4_{j=1}[\alpha_j]_o,\bar{x}+2[\bar{\alpha}_1]_o+2[\bar{\alpha}_3]_o+2[\bar{\alpha}_4]_o)}{\tau(x,\bar{x})}\\ \nonumber
&+\frac{\tau(x,\bar{x}+2[\bar{\alpha}_3]_o)}{\tau(x,\bar{x})}\frac{\dot{\alpha}_{12}}{\ddot{\alpha}_{12}}\frac{\dot{\alpha}_{14}}{\ddot{\alpha}_{14}}\frac{\dot{\alpha}_{24}}{\ddot{\alpha}_{24}}\frac{\tau(x+2\sum^4_{j=1}[\alpha_j]_o,\bar{x}+2[\bar{\alpha}_1]_o+2[\bar{\alpha}_2]_o+2[\bar{\alpha}_4]_o)}{\tau(x,\bar{x})}\\ \nonumber
&-\frac{\tau(x,\bar{x}+2[\bar{\alpha}_4]_o)}{\tau(x,\bar{x})}\frac{\dot{\alpha}_{12}}{\ddot{\alpha}_{12}}\frac{\dot{\alpha}_{13}}{\ddot{\alpha}_{13}}\frac{\dot{\alpha}_{23}}{\ddot{\alpha}_{23}}\frac{\tau(x+2\sum^4_{j=1}[\alpha_j]_o,\bar{x}+2[\bar{\alpha}_1]_o+2[\bar{\alpha}_2]_o+2[\bar{\alpha}_3]_o)}{\tau(x,\bar{x})}
\}.\end{align}
If we shift the variable $x$ as $x-2[\alpha_4]_o$, $\bar{x}$ as $\bar{x}-2[\bar{\alpha}_4]_o$ and remove the $\tau(x,\bar{x})$, we get the fay identity $(1.5)$ in \cite{Chaozhong}. Similar as the proof of the fay identity $(1.5)$\cite{Chaozhong}, we can easily prove that the eq.(\ref{f422}) is equivalent to the two-component BKP hierarchy eq.(\ref{f412}). Based on the two-component BKP hierarchy, we will discuss the reduction of two-component BKP hierarchy.

%\begin{center}
\section{Reduced addition formula of the D type Drinfeld-Sokolov hierarchy }
%\end{center}
\setcounter{section}{5}

Let an integer $n\geq 2$, one gives a constrained condition $L^{2n}=\bar{L}^2=\mathcal{L}$ of the two-component BKP hierarchy. The flows (\ref{f47}),(\ref{f48}) will be influenced by the constrained condition, so we get the reduction of the two-component BKP hierarchy, namely the D type Drinfeld-Sokolov hierarchy.

Firstly, the Lax operator of the D type Drinfeld-Sokolov hierarchy will be achieved by reducing the Lax operator of the two-component BKP hierarchy:\cite{CZ Li}
\begin{equation}
\mathcal{L}=\partial^{2n}+\frac{1}{2}\sum_{i=1}^n\partial^{-1}(v_i\partial^{2i-1}+\partial^{2i-1}v_i)+\partial^{-1}\rho\partial^{-1}\rho.\end{equation}
The dressing structure can be redefined by redefining two fractional operators as
\begin{equation}
\mathcal{L}^{\frac{1}{2n}}=\partial+\sum_{i\geq 1}u_i\partial^{-i},\qquad \mathcal{L}^{\frac{1}{2}}=\partial^{-1}\bar{u}_{-1}+\sum_{i\geq 1}\bar{u}_i\partial^{i}.\end{equation}
Then the dressing structure can be rewritten as the following form:
\begin{equation}
\mathcal{L}^{\frac{1}{2n}}=\Phi \partial\Phi^{-1},\qquad \mathcal{L}^{\frac{1}{2}}=\bar{\Phi}\partial^{-1}\bar{\Phi}^{-1}.\end{equation}
The flows (\ref{f47}),(\ref{f48}) will be influenced by the constrained condition, so the D type Drinfeld-Sokolov hierarchy has the following definition:
\begin{equation}
\frac{\partial\Phi}{\partial t_k}=-(\mathcal{L}^{\frac{k}{2n}})_-\Phi,\qquad \frac{\partial\bar{\Phi}}{\partial t_k}=((\mathcal{L}^{\frac{k}{2n}})_+-\delta_{k1}\mathcal{L}^{-\frac{1}{2}})\bar{\Phi},\end{equation}
\begin{equation}
\frac{\partial\Phi}{\partial \bar{t}_k}=-(\mathcal{L}^{\frac{k}{2}})_-\Phi,\qquad \frac{\partial\bar{\Phi}}{\partial \bar{t}_k}=(\mathcal{L}^{\frac{k}{2}})_+\bar{\Phi},\end{equation}
which $k\in \mathbb{Z}^{odd}_+$.\\
The wave functions will be rewritten as:
\begin{equation}
w(z^{\frac{1}{2n}})=w(t,\bar{t};z^{\frac{1}{2n}})=\Phi e^{\tilde{\xi}(t;z^{\frac{1}{2n}})},\qquad \bar{w}(z^{\frac{1}{2}})=\bar{w}(t,\bar{t};z^{\frac{1}{2}})=\bar{\Phi} e^{xz^{\frac{1}{2}}+\tilde{\xi}(\bar{t};-z^{-\frac{1}{2}})}.\end{equation}

Because of the constrained condition, the bilinear equations (\ref{f411}) can be reduced to the form\cite{SiQi}:
\begin{equation}
res_zz^{2nj-1}w(t',\bar{t'};z)w(t,\bar{t};-z)=res_{\bar{z}}\bar{z}^{-2j-1}\bar{w}(t',\bar{t'};\bar{z})\bar{w}(t,\bar{t};-\bar{z}),\end{equation}
where $j\geq 0$.\\
The bilinear equation for the D type Drinfeld-Sokolov hierarchy can be got
\begin{equation}\label{f58}
\oint\frac{dz}{2\pi i}z^{2nj-1}w(t',\bar{t'};z)w(t,\bar{t};-z)=\oint\frac{d\bar{z}}{2\pi i}\bar{z}^{-2j-1}\bar{w}(t',\bar{t'};\bar{z})\bar{w}(t,\bar{t};-\bar{z}).\end{equation}
The $\tau$-function $\tau_{ds}(t,\bar{t})$ of the D type Drinfeld-Sokolov hierarchy can be got by the bilinear equation:
\begin{align}\label{f59}
\oint\frac{dz}{2\pi i}z^{2nj-1}e^{\tilde{\xi}(t'-t,z)}\tau_{ds}(t'-2[z^{-1}],\bar{t}')\tau_{ds}(t+2[z^{-1}],\bar{t})\\ \nonumber =
\oint\frac{d\bar{z}}{2\pi i}\bar{z}^{-2j-1}e^{\tilde{\xi}(\bar{t}'-\bar{t},\bar{z})}\tau_{ds}(t',\bar{t}'-2[\bar{z}^{-1}])\tau_{ds}(t,\bar{t}+2[\bar{z}^{-1}]).\end{align}

The $\tau_{ds}(t,\bar{t})$ function of the D type Drinfeld-Sokolov hierarchy is defined as eq.(\ref{f59}). Then set $t=x+y$, $t'=x-y$, $\bar{t}=\bar{x}+\bar{y}$, $\bar{t}'=\bar{x}-\bar{y}$, eq.(\ref{f59}) becomes
\begin{align}\label{f510}
\oint\frac{dz}{2\pi i}z^{2nj-1}e^{-2\tilde{\xi}(y,z)}\tau_{ds}(x-y-2[z^{-1}],\bar{x}-\bar{y})\tau_{ds}(x+y+2[z^{-1}],\bar{x}+\bar{y})\\ \nonumber =
\oint\frac{d\bar{z}}{2\pi i}\bar{z}^{-2j-1}e^{-2\tilde{\xi}(\bar{y},\bar{z})}\tau_{ds}(x-y,\bar{x}-\bar{y}-2[\bar{z}^{-1}])\tau_{ds}(x+y,\bar{x}+\bar{y}+2[\bar{z}^{-1}]).\end{align}
Suppose
\begin{equation}
y=\sum^k_{i=1}[\alpha_i]_o,\qquad \bar{y}=\sum^k_{i=1}[\bar{\alpha_i}]_o,\end{equation}
in eq.(\ref{f510}) as Section 4. Then eq.(\ref{f510}) becomes
\begin{align}
\oint\frac{dz}{2\pi i}z^{2nj-1}e^{-2\tilde{\xi}(\sum^k_{i=1}[\alpha_i]_o,z)}\tau_{ds}(x-\sum^k_{i=1}[\alpha_i]_o-2[z^{-1}],\bar{x}-\sum^k_{i=1}
[\bar{\alpha_i}]_o)\\ \nonumber \tau_{ds}(x+\sum^k_{i=1}[\alpha_i]_o+2[z^{-1}],\bar{x}+\sum^k_{i=1}[\bar{\alpha_i}]_o)\\ \nonumber =
\oint\frac{d\bar{z}}{2\pi i}\bar{z}^{-2j-1}e^{-2\tilde{\xi}(\sum^k_{i=1}[\bar{\alpha_i}]_o,\bar{z})}\tau_{ds}(x-\sum^k_{i=1}
[\alpha_i]_o,\bar{x}-\sum^k_{i=1}[\bar{\alpha_i}]_o-2[\bar{z}^{-1}])\\ \nonumber\tau_{ds}(x+\sum^k_{i=1}[\alpha_i]_o,\bar{x}+\sum^k_{i=1}[\bar{\alpha_i}]_o+2[\bar{z}^{-1}]).\end{align}
With (\ref{f417}),(\ref{f418}) and taking residues at $z=\alpha_i^{-1},\bar{z}=\bar{\alpha}_i^{-1}$$(1\leq i\leq {k})$ in the integral, by shifting the variable $x$ as
$
x\rightarrow x+\sum^k_{i=1}[\alpha_i]_o,
\bar{x}\rightarrow \bar{x}+\sum^k_{i=1}[\bar{\alpha}_i]_o,
$ we get the follow theorem.\\
{\sl {\bf  Theorem 4}} The addition formula for the $\tau$-function of the D type Drinfeld-Sokolov hierarchy is:
\begin{align}\label{f513}
\sum^k_{i=1}\prod^k_{j=1,j \neq i}\frac{\alpha_i+\alpha_j}{\alpha_i-\alpha_j}\tau_{ds}(x+2[\alpha_i]_o,\bar{x})
\tau_{ds}(x+2\sum^k_{j=1,j \neq i}[\alpha_j]_o,\bar{x}+2\sum^k_{j=1}[\bar{\alpha}_j]_o)(\alpha_i^{-1})^{2nj} =\\ \nonumber
\sum^k_{i=1} \prod^k_{j=1,j \neq i}\frac{\bar{\alpha}_i+\bar{\alpha}_j}{\bar{\alpha}_i-\bar{\alpha}_j}\tau_{ds}(x,\bar{x}+2[\bar{\alpha}_i]_o) \tau_{ds}(x+2\sum^k_{j=1}[\alpha_j]_o,\bar{x}+2\sum^k_{j=1,j \neq i}[\bar{\alpha}_j]_o)(\bar{\alpha}_i^{-1})^{-2j}.\end{align}
The eq.(\ref{f513}) of the $\tau$-function $\tau_{ds}$ will be simplified into the following equation by dividing $\tau_{ds}(x,\bar{x})^2$.
The addition formula for the $\tau$-function of the D type Drinfeld-Sokolov hierarchy is:
\begin{align}\label{f514}
&\sum^k_{i=1}(-1)^{i-1}\frac{\tau_{ds}(x+2[\alpha_i]_o,\bar{x})}{\tau_{ds}(x,\bar{x})}A^{-1}_{1,\cdots,\hat{i},\cdots,k}\frac{\tau_{ds}(x+2\sum^k_{j=1,j \neq i}[\alpha_j]_o,\bar{x}+2\sum^k_{j=1}[\bar{\alpha}_j]_o)}{\tau_{ds}(x,\bar{x})}(\alpha_i^{-1})^{2nj}\\ \nonumber &=\frac{\bar{A}_{1,\cdots,k}}{A_{1,\cdots,k}}\\ \nonumber
&\times\sum^k_{i=1}(-1)^{i-1}\frac{\tau_{ds}(x,\bar{x}+2[\bar{\alpha}_i]_o)}{\tau_{ds}(x,\bar{x})}\bar{A}^{-1}_{1,\cdots,\hat{i},\cdots,k}\frac{\tau_{ds}(x+2\sum^k_{j=1}[\alpha_j]_o,\bar{x}+2\sum^k_{j=1,j \neq i}[\bar{\alpha}_j]_o)}{\tau_{ds}(x,\bar{x})}(\bar{\alpha}_i^{-1})^{-2j}.\end{align}

In order to show the addition formula of the D type Drinfeld-Sokolov hierarchy clearly, we give the addition formula in the case of $k=4$ as the following example:\\
{\sl {\bf  Example 6}} The case $k=4$ of (\ref{f514}) becomes
\begin{align}\label{f515}
&\frac{1}{\alpha_1^{2nj}}\frac{\tau_{ds}(x+2[\alpha_1]_o,\bar{x})}{\tau_{ds}(x,\bar{x})}\frac{\alpha_{23}}{\tilde{\alpha}_{23}}\frac{\alpha_{24}}{\tilde{\alpha}_{24}}\frac{\alpha_{34}}{\tilde{\alpha}_{34}}\frac{\tau_{ds}(x+2[\alpha_2]_o+2[\alpha_3]_o+2[\alpha_4]_o,\bar{x}+2\sum^4_{j=1}[\bar{\alpha}_j]_o)}{\tau_{ds}(x,\bar{x})}\\ \nonumber
&-\frac{1}{\alpha_2^{2nj}}\frac{\tau_{ds}(x+2[\alpha_2]_o,\bar{x})}{\tau_{ds}(x,\bar{x})}\frac{\alpha_{13}}{\tilde{\alpha}_{13}}\frac{\alpha_{14}}{\tilde{\alpha}_{14}}\frac{\alpha_{34}}{\tilde{\alpha}_{34}}\frac{\tau_{ds}(x+2[\alpha_1]_o+2[\alpha_3]_o+2[\alpha_4]_o,\bar{x}+2\sum^4_{j=1}[\bar{\alpha}_j]_o)}{\tau_{ds}(x,\bar{x})}\\ \nonumber
&+\frac{1}{\alpha_3^{2nj}}\frac{\tau_{ds}(x+2[\alpha_3]_o,\bar{x})}{\tau_{ds}(x,\bar{x})}\frac{\alpha_{12}}{\tilde{\alpha}_{12}}\frac{\alpha_{14}}{\tilde{\alpha}_{14}}\frac{\alpha_{24}}{\tilde{\alpha}_{24}}\frac{\tau_{ds}(x+2[\alpha_1]_o+2[\alpha_2]_o+2[\alpha_4]_o,\bar{x}+2\sum^4_{j=1}[\bar{\alpha}_j]_o)}{\tau_{ds}(x,\bar{x})}\\ \nonumber
&-\frac{1}{\alpha_4^{2nj}}\frac{\tau_{ds}(x+2[\alpha_4]_o,\bar{x})}{\tau_{ds}(x,\bar{x})}\frac{\alpha_{12}}{\tilde{\alpha}_{12}}\frac{\alpha_{13}}{\tilde{\alpha}_{13}}\frac{\alpha_{23}}{\tilde{\alpha}_{23}}\frac{\tau_{ds}(x+2[\alpha_1]_o+2[\alpha_2]_o+2[\alpha_3]_o,\bar{x}+2\sum^4_{j=1}[\bar{\alpha}_j]_o)}{\tau_{ds}(x,\bar{x})}\\ \nonumber
&=\frac{\bar{A}_{1234}}{A_{1234}}\\ \nonumber &\times \{\bar{\alpha}_1^{-2j}\frac{\tau_{ds}(x,\bar{x}+2[\bar{\alpha}_1]_o)}{\tau_{ds}(x,\bar{x})}\frac{\dot{\alpha}_{23}}{\ddot{\alpha}_{23}}\frac{\dot{\alpha}_{24}}{\ddot{\alpha}_{24}}\frac{\dot{\alpha}_{34}}{\ddot{\alpha}_{34}}\frac{\tau_{ds}(x+2\sum^4_{j=1}[\alpha_j]_o,\bar{x}+2[\bar{\alpha}_2]_o+2[\bar{\alpha}_3]_o+2[\bar{\alpha}_4]_o)}{\tau_{ds}(x,\bar{x})}\\ \nonumber
&-\bar{\alpha}_2^{-2j}\frac{\tau_{ds}(x,\bar{x}+2[\bar{\alpha}_2]_o)}{\tau_{ds}(x,\bar{x})}\frac{\dot{\alpha}_{13}}{\ddot{\alpha}_{13}}\frac{\dot{\alpha}_{14}}{\ddot{\alpha}_{14}}\frac{\dot{\alpha}_{34}}{\ddot{\alpha}_{34}}\frac{\tau_{ds}(x+2\sum^4_{j=1}[\alpha_j]_o,\bar{x}+2[\bar{\alpha}_1]_o+2[\bar{\alpha}_3]_o+2[\bar{\alpha}_4]_o)}{\tau_{ds}(x,\bar{x})}\\ \nonumber
&+\bar{\alpha}_3^{-2j}\frac{\tau_{ds}(x,\bar{x}+2[\bar{\alpha}_3]_o)}{\tau_{ds}(x,\bar{x})}\frac{\dot{\alpha}_{12}}{\ddot{\alpha}_{12}}\frac{\dot{\alpha}_{14}}{\ddot{\alpha}_{14}}\frac{\dot{\alpha}_{24}}{\ddot{\alpha}_{24}}\frac{\tau_{ds}(x+2\sum^4_{j=1}[\alpha_j]_o,\bar{x}+2[\bar{\alpha}_1]_o+2[\bar{\alpha}_2]_o+2[\bar{\alpha}_4]_o)}{\tau_{ds}(x,\bar{x})}\\ \nonumber
&-\bar{\alpha}_4^{-2j}\frac{\tau_{ds}(x,\bar{x}+2[\bar{\alpha}_4]_o)}{\tau_{ds}(x,\bar{x})}\frac{\dot{\alpha}_{12}}{\ddot{\alpha}_{12}}\frac{\dot{\alpha}_{13}}{\ddot{\alpha}_{13}}\frac{\dot{\alpha}_{23}}{\ddot{\alpha}_{23}}\frac{\tau_{ds}(x+2\sum^4_{j=1}[\alpha_j]_o,\bar{x}+2[\bar{\alpha}_1]_o+2[\bar{\alpha}_2]_o+2[\bar{\alpha}_3]_o)}{\tau_{ds}(x,\bar{x})}
\}.\end{align}
It is the fact that eq.(\ref{f515}) is equivalent to the D type Drinfeld-Sokolov hierarchy eq.(\ref{f58}).

%\begin{center}
\section{Conclusions and discussions}
%\end{center}

In this paper, we simply introduce the basic knowledge of the discrete KP, the q-deformed KP, the two-component BKP and the D type Drinfeld-Sokolov hierarchies, and we give a detailed derivation of their addition formulae, namely eqs. (\ref{f11})$\thicksim$(\ref{f14}), also we show the equivalence between the hierarchies and the addition formulae. We can know that the addition formulae of other integrable equations can be obtained by the same derivation. So we give the conclusion that the addition formula is of good universality. In our next work, we will begin the study the Giambelli type formula in the above mentioned hierarchies and try to probe more properties about the addition formulae.

%\begin{center}
\section{Acknowledgements}
%\end{center}

 Chuanzhong Li  is supported by  the Zhejiang Provincial Natural Science Foundation under Grant No. LY15A010004, the National Natural Science Foundation of China under Grant No. 11201251, 11571192 and the Natural Science Foundation of Ningbo under Grant No. 2015A610157.
Jingsong He is supported by the National Natural Science Foundation of China under Grant No. 11271210, K. C. Wong Magna Fund in Ningbo University.

%%%%%%%%%%%%%%%%%%%%%%%%%%%%%%%%%%%%%%%%%%%%%%%%%%%%%%%%%%%%%%%%%%%%%%%%%%%%%%%

%%%%%%%%%%%%%%%%% References  %%%%%%%%%%%%%%%%%%%%%%%%%%%%%%%%%%%%%%%
\newpage{}

%%%%%%%%%%%%%%%%%%%%%%%%%%%%%%%%%%%%%%%%%%%%%%%%%%%%%%%%%

\end{document}